\documentclass[%
 reprint,
 amsmath,amssymb,
 aps,
prx,
]{revtex4-2}

\usepackage{graphicx}
\usepackage{dcolumn}
\usepackage{bm}

\usepackage{amsmath, amssymb, amsfonts, amsthm, braket, mathrsfs}

\usepackage{siunitx}
\usepackage[colorlinks, citecolor=blue]{hyperref}

\begin{document}

\preprint{APS/123-QED}

\title{Charge correlator expansion for free fermion negativity}

\author{Yang-Yang Tang}
 \email{tangyy23@mails.tsinghua.edu.cn}
\affiliation{
 Institute for Advanced Study, Tsinghua University, Beijing 100084, China
}

\begin{abstract}

Logarithmic negativity is a widely used entanglement measure
in quantum information theories, which  
can also be efficiently computed in quantum many-body systems
by replica trick or by relating to correlation matrices. In this 
paper, we demonstrate that in free-fermion systems with conserved 
charge, R{\'e}nyi and logarithmic negativity can be expanded 
by connected charge 
correlators, analogous to the case for entanglement 
entropy in the 
context of full counting statistics (FCS). We confirm the rapid 
convergence of this expansion in random all-connected Hamiltonian
through numerical verification, especially for systems with only local
hopping. We find that the replica trick
that get logarithmic negativity from the limit of R{\'e}nyi negativity is
valid in this method only for translational invariant systems. Using this 
expansion, we analyze the scaling behavior of negativity in 
extensive free-fermion systems. In particular, in 1+1 dimensional 
free-fermion systems, we observe that the scaling behavior of 
negativity from our expansion is consistent with known results 
from the method with Toeplitz matrix. These findings provide 
insights into the entanglement properties of free-fermion systems, 
and demonstrate the efficacy of the expansion approach in 
studying entanglement measures.

\end{abstract}

\maketitle


\section{\label{intro}Introduction}

Entanglement, which refers to the non-separability of 
multipartite quantum states
\cite{PhysRevA.40.4277}, is widely regarded as the most 
non-classical phenomenon in quantum mechanics 
\cite{RevModPhys.81.865}. Due to the significant utility of 
entangled states in quantum information theories and experiments, 
various entanglement detections and measures have been 
proposed to detect and quantify the entanglement of quantum 
states \cite{GUHNE20091,10.5555/2011706.2011707}. For pure 
states, the well-known von Neumann entropy uniquely captures 
their entanglement \cite{10.5555/2011706.2011707}. However, for 
mixed states, the condition is complicated, as entanglement entropy 
contains both classical and quantum correlations. It is widely 
believed that there is no unique entanglement measure for mixed 
states, and therefore various entanglement measures are defined 
for different purposes \cite{10.5555/2011706.2011707}.

Among the various entanglement measures, the (logarithmic) negativity, 
also known as partial transpose (PT) negativity
\cite{PhysRevA.65.032314}, is the most 
commonly employed one in the study of quantum many-body systems. 
This measure is particularly useful as it can be computed efficiently 
compared to other measures that rely on variational expressions
\cite{10.5555/2011706.2011707}. 
The negativity measure is inspired by the positive partial 
transpose (PPT) criterion
\cite{PhysRevLett.77.1413,HORODECKI19961}, which claims that for 
separable (non-entangled) states $\rho$, the partial transposition
of the density matrix $\rho^{T_1}$ have no negative eigenvalues. Negativity quantifies 
how negative the eigenvalues of $\rho^{T_1}$ are, and has been 
shown to bound the utility of the state $\rho$ in quantum teleportation
\cite{PhysRevA.65.032314}.
Moreover, the logarithmic negativity is additive for product states 
and thus is extensively used in quantum many-body systems \cite{PhysRevLett.109.130502, Calabrese_2013}.
It has also been 
proven to bound the entanglement that can be asymptotically distilled 
from $\rho$ \cite{PhysRevA.65.032314}.

Fermionic systems possess an alternative definition for negativity, 
known as partial time-reversal (TR) negativity, which exhibits 
superior properties compared to the fermionic PT negativity based 
on the corresponding bosonic density matrix 
\cite{PhysRevB.95.165101,PhysRevLett.118.216402}. The key advantage 
of partial TR negativity is that the partial TR of a fermionic 
Gaussian state remains Gaussian. This feature is particularly 
useful for free-fermion systems, as the partial TR negativity for 
the ground state or thermal equilibrium state at finite temperature 
can then be efficiently computed in terms of two-point correlator 
matrices \cite{PhysRevB.95.165101, Shapourian_2019}. Therefore, 
partial TR negativity offers a more efficient and accurate 
method for studying negativity in fermionic systems.

Negativity can also be computed by replica trick, where the logarithmic
negativity $\mathcal{E}$ is the replica limit $n_e\rightarrow 1$ of R{\'e}nyi
negativity $\mathcal{E}_{n_e}$ \cite{PhysRevLett.109.130502, Calabrese_2013, PhysRevB.95.165101, Shapourian_2019}, 
analogous to the case for
von Neumann entropy and R{\'e}nyi entropy. For certain geometry and dimension,
both bosonic \cite{PhysRevLett.109.130502, Calabrese_2013} and fermionic \cite{PhysRevB.95.165101, Shapourian_2019}
negativity $\mathcal{E}_{n_e}$ can be computed by CFT methods,  
and therefore the scaling behavior of logarithmic
negativity $\mathcal{E}$ can also be analyzed. 

Alternatively, in free-fermion system with conserved charge, we can expand 
these computable entanglement measures into
connected charge correlators by cumulant expansion. It can be proved
that 
$n$-R{\'e}nyi entropy can be expanded by the charge (partial number)
fluctuation as 
\begin{equation}
    \begin{aligned}
        S_A^{(n)}
            =-\frac{1}{1-n}\sum_{\substack{M=2\\
            \mathrm{even}\ M}}^{\infty}&
                2\zeta\left(-M, \frac{n+1}{2}\right)\\
                    &\times\frac{1}{M!}\left(
                        \frac{2\pi i}{n}
                    \right)^M\braket{Q_A^M}_c
    \end{aligned}
\end{equation}
(see Sec. \ref{cumu_entropy} for derivation), where $\zeta(s, a)$ is the generalization of 
the Riemann zeta function known as the Hurwitz zeta function, 
and $\braket{Q_A^M}_c$ is the $M$-th order connected charge correlator
in region $A$. 
Taking the replica limit $n\rightarrow 1$
we get the cumulant expansion for von Neumann entropy \cite{PhysRevX.12.031022}
\begin{equation}\label{vonneumannexpansion}
    \begin{aligned}
        S_A =\sum_{\substack{M=2\\\mathrm{even}\ M}}^{\infty}
            2\zeta(M) \braket{Q_A^M}_c,       
    \end{aligned}
\end{equation}
where $\zeta(s) = \zeta(s, 1)$ is the Riemann zeta function. This result
was equivalently expressed in the context 
of full counting statistics (FCS)
\cite{PhysRevLett.102.100502, PhysRevB.85.035409, Calabrese_2012} as
\begin{equation}
    \begin{aligned}
        S_A =\sum_{\substack{M=2\\\mathrm{even}\ M}}^{\infty}
            (2\pi)^M \frac{|B_M|}{M!}\braket{Q_A^M}_c,
    \end{aligned}
\end{equation}
where $B_M$ are Bernoulli numbers.

Recently, it has been demonstrated
\cite{PhysRevX.12.031022, PhysRevB.109.035413} that 
in $D$-dimensional free-fermion systems with 
Fermi sea, which requires conserved $U(1)$ charge, translational 
symmetry and zero temperature, the $(D+1)$-th order density 
correlator in momentum space is proportional to the Euler 
characteristic $\chi_F$ of the Fermi sea for small $\bm{q}$ 
\begin{equation}\label{univ_corr}
    \int\frac{\mathrm{d}^D\bm{q}_{D+1}}{(2\pi)^D}
    \braket{\rho_{\bm{q}_1}\cdots \rho_{\bm{q}_{D+1}}}_c
    =\frac{\chi_F}{(2\pi)^D}|\det
    \mathbb{Q}|,
\end{equation}
where $\rho_{\bm{q}}=\int {dx}\rho(\bm{x})e^{-i\bm{q}\cdot\bm{x}}$ 
is the charge density operator in momentum space, and $\mathbb{Q}$
is the matrix composed of the momenta $\bm{q}_1,\cdots,\bm{q}_D$.
$\chi_F$ is a topological 
invariant only change at Lifshitz transition. 
Combining \eqref{vonneumannexpansion} and \eqref{univ_corr}, it is proved \cite{PhysRevX.12.031022}
that the multipartite mutual information $\mathcal{I}_{D+1}$ (topological entanglement 
entropy \cite{PhysRevLett.96.110404, PhysRevLett.96.110405})
is proportional to $\chi_F$ of Fermi sea. This proposes
a way to probe the topology of Fermi sea both theoretically
\cite{PhysRevX.12.031022, PhysRevB.109.035413}
and experimentally 
\cite{PhysRevLett.128.076801, PhysRevLett.130.096301, PhysRevB.107.245422},
and further expands our understanding of the entanglement patterns 
related to topology and phase transitions in quantum many-body systems 
\cite{RevModPhys.80.517, 
RevModPhys.82.277, Casini_2009, PhysRevLett.112.160403}.

Inspired by this, we derive the charge correlator 
expansion for partial TR negativity \cite{PhysRevB.95.165101}
\begin{equation}
    \mathcal{E} = \log \mathrm{Tr}\sqrt{\rho^{R_A}
\rho^{R_A\dagger}}.
\end{equation}
in this paper. We briefly summarize the results here and 
leave the definition and derivation in Sec. \ref{cumu_nega}. 
For free-fermion systems with conserved $U(1)$ charge, the cumulant 
expansion for R{\'e}nyi
partial TR negativity \cite{PhysRevB.95.165101} with even $n_e$ 
defined as 
\begin{equation}
    \mathcal{E}_{n_e} = 
\log \mathrm{Tr}(\rho^{R_A}\rho^{R_A\dagger})^{n_e/2}
\end{equation}
is
\begin{equation}
    \begin{aligned}
        \mathcal{E}_{n_e} =& \sum_{\substack{M=2\\\mathrm{even}\ M}}^{\infty}\frac{1}{M!}(\pi i)^M\\
        &\times\left[
        \sum_{p=-\frac{n_e-1}{2}}^{-1/2}
        \left<{\left(-\frac{2p}{n_e}Q_A
        +\left(\frac{2p}{n_e}+1\right)Q_B\right)^M}\right>_c\right.\\
        &\left.
        +\sum_{p=\frac{1}{2}}^{\frac{n_e-1}{2}}
         \left<{\left(-\frac{2p}{n_e}Q_A
        +\left(\frac{2p}{n_e}-1\right)Q_B\right)^M}\right>_c
        \right].
    \end{aligned}
\end{equation}
Written the expansion by orders of $M$ as
\begin{equation}\label{nega_expansion}
    \begin{aligned}
        \mathcal{E}_{n_e} = \sum_{\substack{M=2\\\mathrm{even}\ M}}^{\infty}\mathcal{E}_{n_e}^{(M)},
    \end{aligned}
\end{equation}
by direct calculation we get the first few terms 
\begin{equation}\label{nega_expansion_2}
    \begin{aligned}
        \mathcal{E}_{n_e}^{(2)} =-\pi^2&\left[
        \frac{n_e^2-1}{6n_e}(\braket{Q_A^2}_c+\braket{Q_B^2}_c)\right.\\
        &\left.+\frac{n_e^2+2}{6n_e}\braket{Q_AQ_B}_c
        \right],
    \end{aligned}
\end{equation}
\begin{equation}\label{nega_expansion_4}
    \begin{aligned}
        \mathcal{E}_{n_e}^{(4)} =\pi^4&\left[
        \frac{3n_e^4-10n_e^2+7}{360n_e^3}(\braket{Q_A^4}_c+\braket{Q_B^4}_c)\right.\\
        &+\frac{3n_e^4+10n_e^2-28}{360n_e^3}
        (\braket{Q_A^3Q_B}_c+\braket{Q_AQ_B^3}_c)\\
        &\left.+\frac{n_e^4+14}{120n_e^3}\braket{Q_A^2Q_B^2}_c
        \right]
    \end{aligned}
\end{equation}
and similarly for higher orders. We find that the convergence of the 
expansion depends significantly on the choice of the range of the 
replica momentum $p$ in cumulant expansion (see Sec. \ref{cumu_entropy} and \ref{cumu_nega} for 
details). There is a unique choice for fermionic entropy or negativity to
ensure the best convergence of the expansion, while for bosonic systems there seems to be
no natural way to achieve this. Therefore we suppose that the connected correlator expansion
for computable entanglement measures is limited to fermionic systems.

In Sec. \ref{nume_check}, we numerically check
the validity of this expansion and the quality of convergence 
on thermal equilibrium states of a 100-fermion system with random free Hamiltonian
and conserved charge. The fermionic negativity and charge correlators
are both calculated by the relation with the two-point correlator matrices
for Gaussian states.
We check both Hamiltonian with all-connected hopping
and local hopping to study the effect of locality on the convergence.
The convergence for R{\'e}nyi negativity is generally rapid
even for all-connected free-fermion Hamiltonian, 
and is especially accurate
when there is only local hopping terms, which is the normal
case for extensive
systems. Therefore this expansion can be used to
analyze the scaling behavior of R{\'e}nyi negativity in 
high-dimensional free-fermion systems, especially with local Hamiltonian. 

In the researches about negativity 
\cite{PhysRevLett.109.130502, Calabrese_2013}, 
it is usually assumed that by taking the replica limit 
$n_e\rightarrow 1$ while keeping $n_e$ even we would 
get the logarithmic negativity. 
Taking this limit in \eqref{nega_expansion_2}
and \eqref{nega_expansion_4} we get 
\begin{equation}
    \begin{aligned}
        \mathcal{E}_{n_e\rightarrow 1}^{(2)} =-
        \frac{\pi^2}{2}\braket{Q_AQ_B}_c
    \end{aligned}
\end{equation}
and
\begin{equation}
    \begin{aligned}
        \mathcal{E}_{n_e\rightarrow 1}^{(4)} =\pi^4&\left[
        -\frac{1}{24}
        (\braket{Q_A^3Q_B}_c+\braket{Q_AQ_B^3}_c)\right.\\
        &\left.+\frac{1}{8}\braket{Q_A^2Q_B^2}_c
        \right].
    \end{aligned}
\end{equation}
However, taking the limit $n_e\rightarrow 1$ is
mathematically ambiguous, as arbitrary periodic functions can be 
included
while doing analytic continuation from integers to real numbers.
Therefore it is questionable whether $\mathcal{E}_{n_e\rightarrow 1}=\mathcal{E}$
in a specified method using replica trick. In Sec. \ref{nume_check} we also
check the validity of the expansion for logarithmic negativity. We find that the quality of the expansion
is good only for systems with translational invariant local Hamiltonian.
Fortunately translational invariance is common for quantum many-body systems we study.
Therefore the expansion for $\mathcal{E}_{n_e\rightarrow 1}$
converges to $\mathcal{E}$ and captures its scaling behavior
in translational invariant free-fermion systems with local Hamiltonian, as also confirmed by numerical
results in \cite{PhysRevB.95.165101}.
We leave this as an open question for further
discussion and exploration.

In Sec. \ref{appli} we give 
some applications of \eqref{nega_expansion}
in free-fermion systems of various
dimensions. By applying \eqref{univ_corr} we see 
in (1+1)-d fermion gas (which is also a (1+1)-d CFT system) 
it corresponds with the
result in \cite{PhysRevB.95.165101} from
the method with Toeplitz matrix.

\section{\label{cumu}Correlator expansion for fermion negativity}

\subsection{\label{cumu_entropy}Review: Charge cumulant expansion for fermion entropy}

Charge cumulant expansion for entropy was first derived in the context 
of FCS \cite{PhysRevLett.102.100502}. Here we briefly review the alternative 
derivation by replica method \cite{PhysRevX.12.031022}. The von Neumann entropy 
and $n$-th R{\'e}nyi
entropy of region $A$ in a bipartite system $A\cup B$ is defined as
\begin{equation}
    S_A = -\mathrm{Tr}(\rho_A\log\rho_A)
\end{equation}
and
\begin{equation}
    S_A^{(n)} = \frac{1}{1-n}\log\mathrm{Tr}(\rho_A^n),
\end{equation} 
where $\rho_A = \mathrm{Tr}_B(\rho_{AB})$ is the
reduced density matrix on subsystem $A$.
In replica methods we calculate $\mathrm{Tr}(\rho_A^n)$, and to get von Neumann entropy we take the limit
\begin{equation}
    S_A = \lim_{n\rightarrow 1} S_A^{(n)}
        = \lim_{n\rightarrow 1}\frac{1}{1-n}\log 
        \mathrm{Tr}(\rho_A^n)
\end{equation}
or
\begin{equation}
    S_A = -\left.\frac{\partial \mathrm{Tr}(\rho_A^n)}{\partial n}
        \right|_{n=1}
        = -\left.\frac{\partial\log \mathrm{Tr}(\rho_A^n)}{\partial n}
        \right|_{n=1}.
\end{equation}

$\mathrm{Tr}(\rho_A^n)$ can be written in the form of the expectation
value of a twist operator
by inserting the overcomplete bases of coherent states.
For a single-fermion system,
\begin{equation}
    \mathrm{Tr}(\rho) = \sum_n\braket{n|\rho|n} 
    = \int d\bar{\eta}d\eta e^{-\bar{\eta}\eta}\braket{-\bar{\eta}|\rho|\eta}
\end{equation}
and 
\begin{equation}
    \begin{aligned}
        \mathrm{Tr}(\rho^n) =& \sum_{n_i}\braket{n_1|\rho|n_2}\cdots
        \braket{n_n|\rho|n_1} \\
        =& \int d\bar{\eta}_1d\eta_1\cdots d\bar{\eta}_nd\eta_n
        e^{-\sum_{i}\bar{\eta}_i\eta_i}\\
        &\times\braket{-\bar{\eta}_1|\rho|\eta_2}
        \braket{\bar{\eta}_2|\rho|\eta_3}\cdots
        \braket{\bar{\eta}_n|\rho|\eta_1}\\
        =& \int d\bar{\eta}_1d\eta_1\cdots d\bar{\eta}_nd\eta_n
        e^{-\sum_{i}\bar{\eta}_i\eta_i}\\
        &\times\braket{-\bar{\eta}_1|\rho|-\eta_2}
        \braket{-\bar{\eta}_2|\rho|-\eta_3}\cdots
        \braket{-\bar{\eta}_n|\rho|\eta_1},
    \end{aligned}
\end{equation}
where $\ket{\eta}$ and $\bra{\bar{\eta}}$ are independent coherent 
states of fermion satisfying $\psi\ket{\eta}=\eta\ket{\eta}$
and $\bra{\bar{\eta}}\psi=\bra{\bar{\eta}}\bar{\eta}$, with 
$\eta, \bar{\eta}$ being independent grassmann numbers.
The minus sign comes from the anticommutation relation of 
fermion operator $\psi$. 
Therefore for a bipartite system, we can view $\mathrm{Tr}(\rho_A^n)$ as 
the expectation value of a twist operator $T_{n}$ on $n$
replicas of the original Hilbert space as
\begin{equation}
    \mathrm{Tr}(\rho_A^n) = \braket{T_{n}}, 
\end{equation}
where the operation of $T_{n}$ on the $n$-fold Hilbert space 
in region $A$ is
\begin{equation}
    T_{n}(A)\ket{\eta_1,\cdots, \eta_{n-1},\eta_n}
        = \ket{-\eta_2,\cdots, -\eta_n, \eta_{1}}
\end{equation}
or equivalently for the fermion operators
\begin{equation}
    \begin{aligned}
        T_{n}(A)^{\dagger}\psi_aT_{n}(A)=
        \begin{cases}
            -\psi_{a+1}\ &a<n\\
            \psi_{1}\ &a=n
        \end{cases}
    \end{aligned}.
\end{equation}
We can also write $T_{n}$ in the matrix form as
\begin{equation}\label{twistA}
    \begin{aligned}
        T_{n}(A) = 
        \begin{pmatrix}
            0 & -1 \\
            \ &0 & -1 \\
            \ &\ &\ddots &\ddots\\
            \ &\ &\ &0 & -1 \\
            1 &\ &\ &\ & 0 
        \end{pmatrix}
    \end{aligned}.
\end{equation}
In region $B=\bar{A}$ the operation is the 
identity matrix
\begin{equation}
    \begin{aligned}
        T_{n}(B) = 
        \begin{pmatrix}
            1  \\
            \ &1 \\
            \ &\ &\ddots\\
            \ &\ &\ &1 \\
            \ &\ &\ &\ & 1 
        \end{pmatrix}
    \end{aligned}.
\end{equation}

To calculate $\braket{T_{n}}$, we diagonalize $T_{n}(A)$ 
by introducing Fourier transformation in replica space 
\begin{equation}
    \tilde{\psi}_p(\bm{r}) = \frac{1}{\sqrt{n}}\sum_{a=1}^{n}
        e^{-\frac{2\pi i p}{n}a}
        \psi_a(\bm{r}),
\end{equation}
where
\begin{equation}
    p = -\frac{n-1}{2}, -\frac{n-3}{2}, \cdots, \frac{n-1}{2}
        \ (\mathrm{mod}\ n).
\end{equation}
Here we have the freedom to adjust the choice of the range of $p$ 
by adding integer multiplies of $n$ to any $p$
(\textit{e.g.} change $-(n-1)/2$ to $(n+1)/2$).
In the "momentum" replica space $p$, the operation of $T_{n}$
is
\begin{equation}
    T_{n}^{\dagger}\tilde{\psi}_p(\bm{r})T_{n}
        = \tilde{\psi}_p(\bm{r})\begin{cases}
            -e^{\frac{2\pi i p}{n}}&\bm{r}\in A\\
            1&\bm{r}\in B.
        \end{cases}
\end{equation}
and through the anticommutation relation of fermion operators
we can get
\begin{equation}\label{Tn}
    T_{n} = \prod_{p=-\frac{n-1}{2}}^{\frac{n-1}{2}}
    e^{(\frac{2\pi i p}{n}+\pi i)Q_{A,p}},
\end{equation}
where
\begin{equation}
    Q_{A,p} = \int_{\bm{r}\in A}d^{D}\bm{r}\tilde{\psi}_p^{\dagger}
    (\bm{r})\tilde{\psi}_p(\bm{r})
\end{equation} 
is the total charge operator in $A$ region of the replica $p$.

In some texts there is another form of $T_n(A)$
\begin{equation}\label{twistA_alt}
    \begin{aligned}
        T_{n}(A) = 
        \begin{pmatrix}
            0 & 1 \\
            \ &0 & 1 \\
            \ &\ &\ddots &\ddots\\
            \ &\ &\ &0 & 1 \\
            (-1)^{n+1} &\ &\ &\ & 0 
        \end{pmatrix}
    \end{aligned},
\end{equation}
which will result in 
\begin{equation}\label{Tn2}
    T_{n} = \prod_{p=-\frac{n-1}{2}}^{\frac{n-1}{2}}
    e^{\frac{2\pi i p}{n}Q_{A,p}}.
\end{equation}
For free-fermion systems with conserved $U(1)$ charge (\textit{i.e.} 
with Hamiltonian in the form $H = \sum_{i,j}t_{ij}
c_i^{\dagger}c_j$), \eqref{Tn} and \eqref{Tn2} are equivalent, as
the Hamiltonian for all momentum replica space $p$ are
independent and equals to the original Hamiltonian. 
Thus all $Q_{A,p}=Q_A$, and to get \eqref{Tn2} from \eqref{Tn} 
we just switch the range of $p$. But for interactive systems they may 
differ and possibly have some effects.

As mentioned above, for free-fermion systems with conserved charge,
the momentum replica spaces with different $p$ are independent,
therefore the expectation value of $T_{n}$ decouples into each momentum replica space
\begin{equation}\label{expect_Tn}
    \braket{T_{n}} = \prod_{p}
        \braket{e^{\frac{2\pi i p}{n}Q_{A,p}}}.
\end{equation}
Then we use the cumulant expansion
\begin{equation}
    \log\braket{e^{ikx}} = \exp\left[\sum_{n=1}^{\infty}
    \frac{(ik)^n}{n!}\braket{x^n}_c\right]
\end{equation}
to expand the expectation value and get
\begin{equation}\label{Tnexpanded}
    \begin{aligned}
        \log\braket{T_{n}} &= 
                \sum_p\sum_{M=1}^{\infty}\frac{1}{M!}\left(
                    \frac{2\pi i p}{n}
                \right)^M\braket{Q_A^M}_c\\
            &=\sum_{M=1}^{\infty}
                \left(\sum_p p^M\right)\frac{1}{M!}\left(
                    \frac{2\pi i}{n}
                \right)^M\braket{Q_A^M}_c.
    \end{aligned}
\end{equation}
For the range of $p$ in summation $\sum_p p^M$, we choose 
\begin{equation}\label{prange2}
    p = -\frac{n-1}{2}, -\frac{n-3}{2}, \cdots, \frac{n-1}{2}.
\end{equation}
Note that this is the only natural choice to make $\sum_p p^M$ smallest
for even $M$ and 0 for odd $M$. 
Although the choice of range have no effect for
\eqref{expect_Tn} as the eigenvalues of charge operator $Q_{A,p}$ are all integers,
it matters for \eqref{Tnexpanded}.
A different choice from \eqref{prange2} will lead to nonzero 
imaginary odd-$M$ terms in \eqref{Tnexpanded}, and the convergence
for even-$M$ terms will also become bad. 
After truncation to a finite order of $M$, it will fail to give 
the correct coefficient, especially when we use the expansion to
analyze the scaling behavior of entropy in extensive systems.
This comes from
the multivaluedness of log function, analogous to the 
expansion for 1 as
\begin{equation}
    \begin{aligned}
        1 = e^{2\pi i N}
        = \sum_{M=1}^{\infty}\frac{1}{M!}(2\pi i N)^M
    \end{aligned}
\end{equation}
where $N$ is an integer, which is also bad for $N\neq 0$.

\eqref{Tnexpanded} and \eqref{prange2} gives the cumulant expansion for R{\'e}nyi entropy
\begin{equation}\label{expan_renyi}
    \begin{aligned}
        S_A^{(n)}
            =-\frac{1}{1-n}\sum_{\substack{M=2\\
            \mathrm{even}\ M}}^{\infty}&
                2\zeta\left(-M, \frac{n+1}{2}\right)\\
                    &\times\frac{1}{M!}\left(
                        \frac{2\pi i}{n}
                    \right)^M\braket{Q_A^M}_c
    \end{aligned}
\end{equation}
(see Appendix \ref{appendix_entropy} for derivation). 
By taking limit $n\rightarrow 1$ we get the cumulant expansion for von Neumann entropy
\begin{equation}\label{expan_vonneumann}
    \begin{aligned}
        S_A =\sum_{\substack{M=2\\\mathrm{even}\ M}}^{\infty}
            2\zeta(M) \braket{Q_A^M}_c   
    \end{aligned}
\end{equation}
(see Appendix \ref{appendix_entropy} or \cite{PhysRevX.12.031022} for derivation),
where $\zeta(M)$ is the Riemann zeta function.

\subsection{\label{cumu_nega}Charge correlator expansion for 
fermion negativity}

For a bipartite bosonic system $A\cup B$, (logarithmic) PT negativity is defined as 
\begin{equation}
    \mathcal{E} = \log ||\rho^{T_A}||_1 = 
    \log \mathrm{Tr}\sqrt{(\rho^{T_A})^2}.
\end{equation}
where $\rho^{T_A}$ is
the partial transposition of the density matrix $\rho$ 
on the subsystem $A$, defined 
in the occupation number basis as
\begin{equation}
    \begin{aligned}
    (\ket{n_A}&\otimes\ket{n_B}\bra{m_A}\otimes\bra{m_B})^{T_A}\\
    &=\ket{m_A}\otimes\ket{n_B}\bra{n_A}\otimes\bra{m_B}.
    \end{aligned}
\end{equation}

For fermionic systems, we use the alternative definition \cite{PhysRevB.95.165101} called
partial TR negativity
\begin{equation}\label{nega_defi}
    \mathcal{E} = \log ||\rho^{R_A}||_1 = 
    \log \mathrm{Tr}\sqrt{\rho^{R_A}\rho^{R_A\dagger}},
\end{equation}
where $\rho^{R_A}$ is the partial time-reversal of $\rho_{AB}$ on 
subsystem $A$, defined in the coherent basis as
\begin{equation}
    U(\ket{\eta}\bra{\bar{\eta}})^RU^{\dagger}
    =\ket{i\bar{\eta}}\bra{i\eta},
\end{equation}
where $U$ is the unitary operator in time-reversal operator
and have no effect in the trace.
In the occupation number basis it is 
\begin{equation}
    \begin{aligned}
    U(\ket{n_A}&\otimes\ket{n_B}\bra{m_A}\otimes\bra{m_B}
    )^{R_A}U^{\dagger}\\
    &=(-1)^{\phi(\{n\},\{m\})}
    \ket{m_A}\otimes\ket{n_B}\bra{n_A}\otimes\bra{m_B},
    \end{aligned}
\end{equation}
where the phase factor is
\begin{equation}
    \begin{aligned}
        \phi(\{n\},\{m\}) =& \frac{\tau_1(\tau_1+2)}{2}
        +\frac{\bar{\tau}_1(\bar{\tau}_1+2)}{2} + \tau_2\bar{\tau}_2 \\&+ \tau_1\tau_2 +\bar{\tau}_1\bar{\tau}_2 +(\bar{\tau}_1+\bar{\tau}_2)
        (\tau_1+\tau_2),
    \end{aligned}
\end{equation}
and $\tau_{1(2)} = \sum_{i\in A(B)}n_i$, $\bar{\tau}_{1(2)} = \sum_{i\in A(B)}m_i$.

In the following we consider a fermionic field system devided into three 
parts $A$, $B$ and $C$, and still use \eqref{nega_defi}
to define the negativity 
of $A$ and $B$ by $\rho_{AB}=\mathrm{Tr}_C\rho_{ABC}$ in order to
measure the entanglement between region $A$ and $B$. 
We can also use replica method \cite{PhysRevLett.109.130502, 
Calabrese_2013, PhysRevB.95.165101} to calculate negativity.
The $n$-th R{\'e}nyi negativity is defined as \cite{PhysRevB.95.165101} 
\begin{equation}\label{defi_renyi_nega}
    \mathcal{E}_{n} = 
    \begin{cases}
        \log \mathrm{Tr}\left(\rho^{R_A}\rho^{R_A\dagger}\cdots
        \rho^{R_A}\rho^{R_A\dagger}\right)
        \ &n\ \mathrm{even} \\
        \log \mathrm{Tr}\left(\rho^{R_A}\rho^{R_A\dagger}\cdots
        \rho^{R_A}\right)
        \ &n\ \mathrm{odd} 
    \end{cases}
\end{equation}
and it is usually assumed that by taking limit $n_e\rightarrow 1$
while keeping $n_e$ even we would get the negativity $\mathcal{E}$.
In \cite{PhysRevB.95.165101} it is proved that R{\'e}nyi negativity
can also be written as the expectation value of a twist operator
\begin{equation}
    \mathcal{E}_{n} = \braket{T_n},
\end{equation}
where the operation of the twist operator on the three regions is
\begin{equation}\label{twistA1}
    \begin{aligned}
        T_{n}(A) = 
        \begin{pmatrix}
            0 &\ &\ &\ &(-1)^{n-1} \\
            1 &0 \\
            \ &1 & 0 \\
            \ &\ &\ddots &\ddots \\
            &\ &\ &1 &0 
        \end{pmatrix}
    \end{aligned},
\end{equation}
\begin{equation}\label{twistA2}
    \begin{aligned}
        T_{n}(B) = 
        \begin{pmatrix}
            0 & -1 \\
            \ &0 & -1 \\
            \ &\ &\ddots &\ddots\\
            \ &\ &\ &0 & -1 \\
            1 &\ &\ &\ & 0 
        \end{pmatrix}
    \end{aligned},
\end{equation}
and 
\begin{equation}
    \begin{aligned}
        T_{n}(C) = 
        \begin{pmatrix}
            1  \\
            \ &1 \\
            \ &\ &\ddots\\
            \ &\ &\ &1 \\
            \ &\ &\ &\ & 1 
        \end{pmatrix}
    \end{aligned}.
\end{equation}
The matrices \eqref{twistA1} and \eqref{twistA2} can be 
simultaneously diagonalized only when $n$ is even.
Similarly with
\eqref{Tn}, for even $n_e$ it is direct to get
\begin{equation}
    T_{n_e} = \prod_{p=-\frac{n_e-1}{2}}^{\frac{n_e-1}{2}}
    e^{-\frac{2\pi i p}{n_e}Q_{A,p}
    +(\frac{2\pi i p}{n_e}+\pi i)Q_{B,p}},
\end{equation}
which is equivalent to (57) of \cite{PhysRevB.95.165101}
for free-fermion systems with conserved $U(1)$ charge,
since they differ with only a shift of the range of $p$.
Similarly with the case for R{\'e}nyi entropy, 
for free-fermion systems with conserved charge
the expectation value of $T_{n}$ decouples into each momentum replica space
\begin{equation}\label{expect_Tne}
    \braket{T_{n_e}} = \prod_{p=-\frac{n_e-1}{2}}^{\frac{n_e-1}{2}}
        \braket{e^{-\frac{2\pi i p}{n_e}Q_{A,p}
    +(\frac{2\pi i p}{n_e}+\pi i)Q_{B,p}}}.
\end{equation}

Now in order to get the cumulant expansion, 
we shift the phase before $Q_{B,p}$ by $2\pi i$ for the positive 
half of $p$, so
\begin{equation}
    \begin{aligned}
         \braket{T_{n_e}} = 
         &\prod_{p=-\frac{n_e-1}{2}}^{-\frac{1}{2}}
            \braket{e^{-\frac{2\pi i p}{n_e}Q_{A,p}
        +(\frac{2\pi i p}{n_e}+\pi i)Q_{B,p}}}\\
        &\times\prod_{p=\frac{1}{2}}^{\frac{n_e-1}{2}}
            \braket{e^{-\frac{2\pi i p}{n_e}Q_{A,p}
        +(\frac{2\pi i p}{n_e}-\pi i)Q_{B,p}}}.
    \end{aligned}
\end{equation}
Then we apply the cumulant expansion and get
\begin{equation}\label{nega_expansion_pre}
    \begin{aligned}
        \mathcal{E}_{n_e} =& \sum_{M=1}^{\infty}\frac{1}{M!}(\pi i)^M\\
        &\times\left[
        \sum_{p=-\frac{n_e-1}{2}}^{-1/2}
        \left<{\left(-\frac{2p}{n_e}Q_A
        +\left(\frac{2p}{n_e}+1\right)Q_B\right)^M}\right>_c\right.\\
        &\left.
        +\sum_{p=\frac{1}{2}}^{\frac{n_e-1}{2}}
         \left<{\left(-\frac{2p}{n_e}Q_A
        +\left(\frac{2p}{n_e}-1\right)Q_B\right)^M}\right>_c
        \right].
    \end{aligned}
\end{equation}
Here the range of $p$ is chosen for a similar reason
as \eqref{Tnexpanded}. This is also the only way to make 
the coefficient
smallest for even $M$ and 0 for odd $M$,
which is the key point for the convergence of the expansion. 
Written the expansion by orders of $M$ as
\begin{equation}\label{nega_expansion_sec}
    \begin{aligned}
        \mathcal{E}_{n_e} = \sum_{\substack{M=2\\\mathrm{even}\ M}}^{\infty}\mathcal{E}_{n_e}^{(M)},
    \end{aligned}
\end{equation}
Summing over $p$ we get the first two terms
\begin{equation}\label{nega_expansion_2_sec}
    \begin{aligned}
        \mathcal{E}_{n_e}^{(2)} =-\pi^2&\left[
        \frac{n_e^2-1}{6n_e}(\braket{Q_A^2}_c+\braket{Q_B^2}_c)\right.\\
        &\left.+\frac{n_e^2+2}{6n_e}\braket{Q_AQ_B}_c
        \right],
    \end{aligned}
\end{equation}
and
\begin{equation}\label{nega_expansion_4_sec}
    \begin{aligned}
        \mathcal{E}_{n_e}^{(4)} =\pi^4&\left[
        \frac{3n_e^4-10n_e^2+7}{360n_e^3}(\braket{Q_A^4}_c+\braket{Q_B^4}_c)\right.\\
        &+\frac{3n_e^4+10n_e^2-28}{360n_e^3}
        (\braket{Q_A^3Q_B}_c+\braket{Q_AQ_B^3}_c)\\
        &\left.+\frac{n_e^4+14}{120n_e^3}\braket{Q_A^2Q_B^2}_c
        \right].
    \end{aligned}
\end{equation}
It is also direct for higher orders. We have numerically 
checked that this expansion converges rapidly for thermal equilibrium 
states of randomly 
generated all-connected free Hamiltonian, see Sec. \ref{nume_check}.
For Hamiltonian with local random hopping, the expansion is extremely 
accurate at low temperature. 

Taking the limit $n_e\rightarrow 1$ we get 
\begin{equation}\label{log_nega_2}
    \begin{aligned}
        \mathcal{E}_{n_e\rightarrow 1}^{(2)} 
        =-\frac{\pi^2}{2}\braket{Q_AQ_B}_c
    \end{aligned}
\end{equation}
and
\begin{equation}\label{log_nega_4}
    \begin{aligned}
        \mathcal{E}_{n_e\rightarrow 1}^{(4)} =\pi^4&\left[
        -\frac{1}{24}
        (\braket{Q_A^3Q_B}_c+\braket{Q_AQ_B^3}_c)\right.\\
        &\left.+\frac{1}{8}\braket{Q_A^2Q_B^2}_c
        \right].
    \end{aligned}
\end{equation}
The analytic continuation is ambiguous
mathematically, and it is questionable whether $\mathcal{E}_{n_e
\rightarrow 1}=\mathcal{E}$. In Sec. \ref{nume_check} we find that 
the correspondence of the expansion for $\mathcal{E}_{n_e
\rightarrow 1}$ with $\mathcal{E}$ requires translational
invariance in the Hamiltonian.

\section{\label{nume_check}Numerical verification}

\begin{figure*}
    \centering
    \includegraphics[scale=0.33]{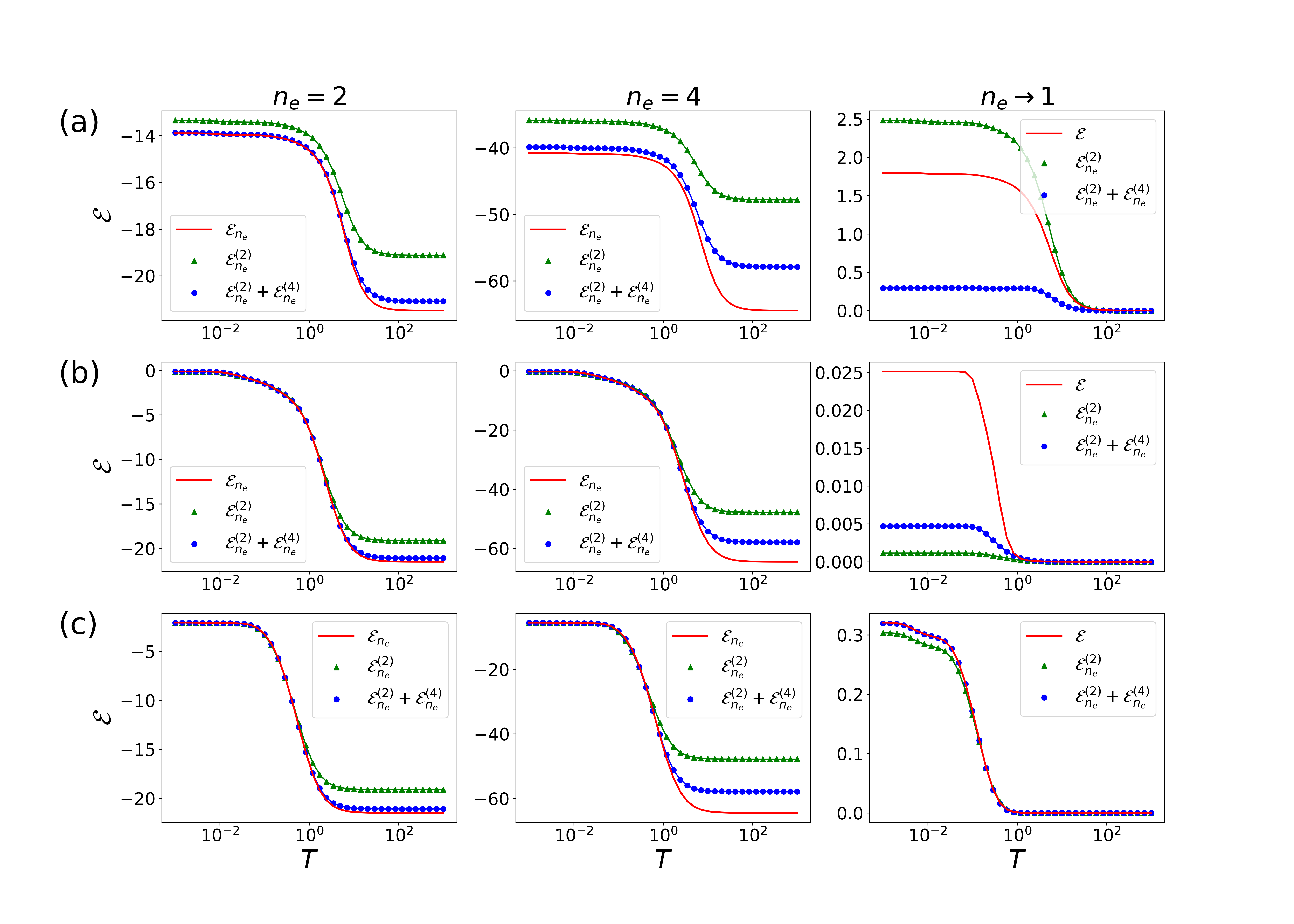}
    \caption{Numerical verification for the validity of charge correlator expansion
    \eqref{nega_expansion_pre} for $n_e=2,4$ and $n_e\rightarrow 1$ 
    on thermal equilibrium states 
    of 100-fermion system with conserved charge, \textit{i.e.}
    $\rho=e^{-\beta H}/Z$, $\beta=1/T$ and $H=\sum_{i,j}
    t_{ij}c_i^{\dagger}c_j$. The amplitude of each $t_{ij}$ 
    is randomly generated from a normal distribution, and the phase
    is generated from an even distribution in $[0, 2\pi)$. 
    The first two terms of the expansion \eqref{nega_expansion_2_sec} 
    and \eqref{nega_expansion_4_sec} is calculated by \eqref{eqns_first} to \eqref{eqns_middle}, and the
    exact values of negativity are calculated by \eqref{eqns_final1} and \eqref{eqns_final2}.
    The regions $A$ and $B$ are chosen to be the 21 to 30-th and 34 to 54-th fermions respectively.
    The quality of convergence depends on the pattern of the random
    Hamiltonian generated.
    (a) With random all-connected hopping $t_{ij}$, generated from the same distribution between all sites.
    (b) With random local hopping $t_{ij}$ for 1+1-d system. The amplitude of $t_{ij}$
    exponentially decays with the distance $|i-j|$.
    (c) With random translational invariant local hopping $t_{ij} = t_{i-j, 0}$ for 1+1-d system,
    within the distance of 3 atoms $|i-j|\leq 3$.}
    \label{fig_pttest}
\end{figure*}

To check the validity of \eqref{nega_expansion_pre}, we numerically calculate 
the expansion \eqref{nega_expansion_2_sec}, \eqref{nega_expansion_4_sec}, \eqref{log_nega_2} and
\eqref{log_nega_4} in 
thermal equilibrium states of a 100-fermion system with randomly generated free Hamiltonian 
and conserved charge. 
It is well known that R{\'e}nyi and von Neumann entropy of Gaussian states can be computed
efficiently by relating to the two-point correlator matrices \cite{PhysRevB.64.064412, IngoPeschel_2003, PhysRevB.69.075112}.
Both R{\'e}nyi and logarithmic negativity of Gaussian states
can also be related to correlator matrices, and so do
the connected charge correlators.
The correlator matrix of a bipartite system $A\cup B$ can be partitioned to block
matrices as 
\begin{equation}
    C = \begin{pmatrix}
        C^{11}&C^{12}\\
        C^{21}&C^{22}
    \end{pmatrix},
\end{equation}
where the matrix elements are $C_{ij}=\braket{c_i^{\dagger}c_j}$. $C^{11}$ and
$C^{22}$ are the reduced correlator matrices on the subsystems $A$ and $B$, and
$C^{12}$ and $C^{21}$ contain correlation between two regions.
Using Wick's theorem, the connected charge correlators in the expansion can be reduced to the
trace of the product of the block matrices as
\begin{equation}\label{eqns_first}
    \braket{Q_A^2}_c = \mathrm{Tr}C^{11} - \mathrm{Tr}(C^{11})^2,
\end{equation}
\begin{equation}
    \braket{Q_AQ_B}_c =  - \mathrm{Tr}(C^{12}C^{21}),
\end{equation}
\begin{equation}
    \braket{Q_B^2}_c = \mathrm{Tr}C^{22} - \mathrm{Tr}(C^{22})^2,
\end{equation}
\begin{equation}
    \braket{Q_A^4}_c = \mathrm{Tr}C^{11} - 7\mathrm{Tr}(C^{11})^2 
    + 12\mathrm{Tr}(C^{11})^3- 6\mathrm{Tr}(C^{11})^4,
\end{equation}
\begin{equation}
    \begin{aligned}
        \braket{Q_A^3Q_B}_c = & - \mathrm{Tr}(C^{12}C^{21})
        + 6\mathrm{Tr}(C^{11}C^{12}C^{21}) \\ &- 6\mathrm{Tr}(C^{11}C^{11}C^{12}C^{21}),
    \end{aligned}
\end{equation}
\begin{equation}
    \begin{aligned}
        \braket{Q_A^2Q_B^2}_c =&  - \mathrm{Tr}(C^{12}C^{21})
        + 2\mathrm{Tr}(C^{11}C^{12}C^{21}) \\ &+ 2\mathrm{Tr}(C^{22}C^{21}C^{12}) 
        - 4\mathrm{Tr}(C^{11}C^{12}C^{21}C^{12}) \\ &- 2 \mathrm{Tr}(C^{12}C^{21})^2,
    \end{aligned}
\end{equation}
\begin{equation}
    \begin{aligned}
        \braket{Q_AQ_B^3}_c =&  - \mathrm{Tr}(C^{12}C^{21})
        + 6\mathrm{Tr}(C^{22}C^{21}C^{12}) \\ &- 6\mathrm{Tr}(C^{22}C^{22}C^{21}C^{12}),
    \end{aligned}
\end{equation}
\begin{equation}\label{eqns_middle}
    \braket{Q_B^4}_c = \mathrm{Tr}C^{22} - 7\mathrm{Tr}(C^{22})^2 
    + 12\mathrm{Tr}(C^{22})^3- 6\mathrm{Tr}(C^{22})^4.
\end{equation}

To calculate the negativity of Gaussian states, we follow the method developed in 
\cite{PhysRevB.95.165101} and \cite{Shapourian_2019}. Define the covariance matrices
\begin{equation}
    \Gamma = \mathbb{I} - 2C = \begin{pmatrix}
        \Gamma^{11}&\Gamma^{12}\\
        \Gamma^{21}&\Gamma^{22}
    \end{pmatrix}
\end{equation}
(where $\mathbb{I}$ is the identity matrix), the corresponding transformed matrices
\begin{equation}
    \Gamma_{\pm}  = \begin{pmatrix}
        -\Gamma^{11}&\pm i\Gamma^{12}\\
        \pm i\Gamma^{21}&\Gamma^{22}
    \end{pmatrix},
\end{equation}
and a new single particle correlation function 
\begin{equation}
    C_{\Xi} = \frac{1}{2}[\mathbb{I} - (\mathbb{I}+\Gamma_+\Gamma_-)^{-1}(\Gamma_++\Gamma_-)].
\end{equation}
Denoting $\zeta_i$ and $\xi_i$ as the eigenvalues of $C$ and $C_{\Xi}$ respectively, 
the logarithmic and R{\'e}nyi negativity for free system with conserved charge is
\begin{equation}\label{eqns_final1}
    \mathcal{E} = \sum_j \log[\xi_j^{1/2}+(1-\xi_j)^{1/2}] 
        + \frac{1}{2}\sum_j \log[\zeta_j^2+(1-\zeta_j)^2]
\end{equation}
and
\begin{equation}\label{eqns_final2}
    \begin{aligned}
        \mathcal{E}_{n_e} =& \sum_j \log[\xi_j^{n_e/2}+(1-\xi_j)^{n_e/2}] 
            \\ &+ \frac{n_e}{2}\sum_j \log[\zeta_j^2+(1-\zeta_j)^2].
    \end{aligned}
\end{equation}

Using \eqref{eqns_first} to \eqref{eqns_middle}, we calculate the
first two orders of the charge correlator expansion \eqref{nega_expansion_2_sec} 
and \eqref{nega_expansion_4_sec}
and compare it
with the exact $\mathcal{E}$ and $\mathcal{E}_{n_e}$ calculated by 
\eqref{eqns_final1} and \eqref{eqns_final2}, see 
Fig. \ref{fig_pttest}. From Fig. \ref{fig_pttest} (a)(b) we see that the convergence of R{\'e}nyi negativity
is generally fast even for all-connected Hamiltonian. When there are only local hopping terms
that decays exponentially with the distance,
which is the normal case for extensive systems, the convergence is extremely accurate
at low temperature. 
For high temperature the convergence becomes slow,  
particularly for higher $n_e$. 
Note that these phenomena are all similar to the case of the expansion for R{\'e}nyi and
von Neumann entropy \eqref{expan_renyi} and \eqref{expan_vonneumann}, where the convergence
is better for systems with low temperature and local hopping terms.

However, the correspondence of the expansion at replica limit $n_e\rightarrow 1$ with logarithmic negativity 
requires more constraint.
Fig.\ref{fig_pttest} (b) shows the bad convergence even for local random Hamiltonian,
from Fig.\ref{fig_pttest} (c) we see that the convergence 
is valid for translational invariant systems. With translational symmetry, the convergence is also
better in systems with local hopping terms.

Therefore, from numerical verification we conclude that the charge correlator
expansion can be used to analyze the scaling behavior of R{\'e}nyi negativity for local free
Hamiltonian with conserved charge, while for logarithmic negativity it further requires
translational invariance. In \cite{PhysRevB.95.165101} it is numerically verified that 
the scaling behavior of negativity in Kitaev chain 
and SSH model (both with translational invariant Hamiltonian)
is consistent with the CFT result from replica trick.

\section{\label{appli}Applications}

\subsection{\label{1dfree}(1+1)-d free fermion}
(1+1)-d free-fermion systems with translational symmetry at zero temperature, with the Euler characteristic of the Fermi sea being $\chi_F$ (equals the number of segments of the Fermi sea), is also a CFT with central charge 
$c=\chi_F$ \cite{PhysRevX.12.031022}. For (1+1)-d 
free-fermion systems, 
the $M=2$ term is related to $\chi_F$
and \eqref{univ_corr} becomes the universal second-order density correlator
\begin{equation}
    \int \frac{dq'}{2\pi}\braket{\rho_q\rho_{q'}}_c
    =\chi_F |q|,
\end{equation}
for small $q$. where $\rho_q$ is the charge density in momentum space
\begin{equation}
    \rho_q= \int {dx}\rho(x)e^{-iqx}
    =\int \frac{dk}{2\pi}c_k^{\dagger}c_{k+q}.
\end{equation}
After Fourier transformation we get density correlator $\braket{\rho(x_1)\rho(x_2)}_c$ in real space.
Here we have to apply the invariance of correlator under translation in systems with translational symmetry by
\begin{equation}\label{corr_real}
    \begin{aligned}
        \braket{\rho(x_1)\rho(x_2)}_c&=\braket{\rho(x_1-x_2)\rho(0)}_c\\
        &=\int \frac{dq}{2\pi}\frac{dq'}{2\pi}\braket{\rho_q\rho_{q'}}_ce^{iq(x_1-x_2)}\\
        &=-\frac{\chi_F}{2\pi^2}\frac{1}{(x_1-x_2)^2}.
    \end{aligned}
\end{equation}
\eqref{corr_real} applies for large $x_1-x_2\gg \varepsilon $, 
where the short-distance cutoff $\varepsilon $ is about
the inverse of the curvature of Fermi surface \cite{PhysRevX.12.031022} (in 
1d the length of the shortest segment or interval of Fermi sea).

To get charge correlator of regions we integrate \eqref{corr_real} in real space.
Now suppose we take the region $A:\ -l_1\leq x\leq 0$, 
$B:\ d\leq x\leq d+l_2$ and $C$ being the rest parts. 
For $\braket{Q_A^2}_c$ we encounter the case for short distance
with small $x_1-x_2$. To solve this problem we consider 
a simplified model, where for $|x_1-x_2|>\varepsilon$ 
\begin{equation}
    \braket{\rho(x_1)\rho(x_2)}_c
    =-\frac{\chi_F}{2\pi^2}\frac{1}{(x_1-x_2)^2},
\end{equation}
and for $|x_1-x_2|\leq\varepsilon$ it is a constant
\begin{equation}
    \braket{\rho(x_1)\rho(x_2)}_c
    =-\frac{\chi_F}{2\pi^2}\frac{1}{\varepsilon^2}.
\end{equation}
We consider the case that the regions $A,B$ are much larger 
than the short-distance cutoff,  
\textit{i.e.} $l_1, l_2\gg \varepsilon $.
Therefore 
\begin{equation}
    \begin{aligned}
        \braket{Q_A^2}_c 
        &= \int_{-l_1}^{0}dx_1\int_{-l_1}^{0}dx_2
        \braket{\rho(x_1)\rho(x_2)}_c\\
        &= \frac{\chi_F}{2\pi^2}\left[2\log\frac{l_1}
        {\varepsilon}
        +O\left(\frac{l_1}
        {\varepsilon}\right)\right].
    \end{aligned}
\end{equation}
Similarly for
$\braket{Q_B^2}_c$ we have
\begin{equation}
    \begin{aligned}
        \braket{Q_B^2}_c 
        &= \frac{\chi_F}{2\pi^2}\left[2\log\frac{l_2}
        {\varepsilon}
        +O\left(\frac{l_2}
        {\varepsilon}\right)\right].
    \end{aligned}
\end{equation}
For $\braket{Q_AQ_B}_c$, at $d\geq\varepsilon$ the integration get 
\begin{equation}
    \begin{aligned}
        \braket{Q_AQ_B}_c 
        &= \int_{-l_1}^{0}dx_1\int_{d}^{d+l_2}dx_2
        \braket{\rho(x_1)\rho(x_2)}_c\\
        &= -\frac{\chi_F}{2\pi^2}\log
            \frac{(l_1+d)(l_2+d)}{d(l_1+l_2+d)}
    \end{aligned}
\end{equation}
and for $d<\varepsilon\ll l_1,l_2$,
\begin{equation}
    \begin{aligned}
        \braket{Q_AQ_B}_c 
        &= \int_{-l_1}^{0}dx_1\int_{d}^{d+l_2}dx_2
        \braket{\rho(x_1)\rho(x_2)}_c\\
        &= -\frac{\chi_F}{2\pi^2}\left[\log
            \frac{(l_1+d)(l_2+d)}{\varepsilon(l_1+l_2+d)}
            +O\left(\frac{d}{\varepsilon}\right)
            \right]\\
        &= -\frac{\chi_F}{2\pi^2}\log
        \frac{l_1l_2}{\varepsilon(l_1+l_2)}.
    \end{aligned}
\end{equation}
In the last step we ignore the small $O(d/\varepsilon)$
terms.
With the above results for charge correlators, 
for $d\gg\varepsilon$
\eqref{nega_expansion_2_sec} becomes
\begin{equation}\label{1dfree_2}
    \begin{aligned}
        \mathcal{E}_{n_e}^{(2)} =&\chi_F\left[
        -\frac{n_e^2-1}{6n_e}\left(
        \log\frac{l_1l_2}{\varepsilon^2}
        +O\left(\frac{l}{\varepsilon}\right)\right)\right.\\
        &+\left.\frac{n_e^2+2}{12n_e}
        \log\frac{(l_1+d)(l_2+d)}{d(l_1+l_2+d)}\right].
    \end{aligned}
\end{equation}
And for $d\rightarrow 0$ it becomes
\begin{equation}\label{1dfree_2_2}
    \begin{aligned}
        \mathcal{E}_{n_e}^{(2)} =&\chi_F\left[
        -\frac{n_e^2-1}{6n_e}\left(
        \log\frac{l_1l_2}{\varepsilon^2}
        +O\left(\frac{l}{\varepsilon}\right)\right)\right.\\
        &+\left.\frac{n_e^2+2}{12n_e}\log
        \frac{l_1l_2}{\varepsilon(l_1+l_2)}\right].
    \end{aligned}
\end{equation}

First we consider the case $d\rightarrow 0$,
\textit{i.e.} regions $A$ and $B$ are adjacent,
\eqref{1dfree_2_2} can also be written as
\begin{equation}
    \begin{aligned}
        \mathcal{E}_{n_e}^{(2)} = \chi_F&\left[
        -\frac{n_e^2-4}{12n_e}\log\frac{l_1l_2}
        {\varepsilon^2}-\frac{n_e^2+2}{12n_e}
        \log\frac{l_1+l_2}{\varepsilon^2}\right.\\
        &\left.
        -\frac{n_e^2-1}{6n_e}
        O\left(\frac{l}{\varepsilon}\right)
        \right],
    \end{aligned}
\end{equation}
which is identical to (85) in \cite{PhysRevB.95.165101} 
at even $n_e$ except for the short distance
cutoff (note that the term for short distance
cutoff
$O\left(l/{\varepsilon}\right)$
is generally large and cannot be neglected). Taking the limit $n_e\rightarrow 1$
we get
\begin{equation}\label{nega_n1_1d}
    \begin{aligned}
        \mathcal{E}_{n_e\rightarrow 1}^{(2)}=
        \frac{\chi_F}{4}\log\frac{l_1l_2}{\varepsilon(l_1+l_2)}.
    \end{aligned}
\end{equation}
Then the first term related to 
divergent short-distance cutoff
is eliminated, which is consistent with what we expect 
for negativity.
Numerical checks in Sec. \ref{nume_check} have confirmed that for system with translational symmetry, 
the expansion of $\mathcal{E}_{n_e\rightarrow 1}$ converges to $\mathcal{E}$, 
so \eqref{nega_n1_1d} captures the scaling property of $\mathcal{E}$.

We can also consider the cases that region $A$ and $B$ are far apart,
\textit{i.e.} $d\gg l_1,l_2$. Then \eqref{1dfree_2} can be 
expanded to the second order of $l_i/d$ as
\begin{equation}
    \begin{aligned}
        \mathcal{E}_{n_e}^{(2)} = \chi_F&\left[
        -\frac{n_e^2-1}{6n_e}\left(
            \log\frac{l_1l_2}{\varepsilon^2}
            +O\left(\frac{l}{\varepsilon}\right)\right)\right. \\
        &+\left.\frac{n_e^2+2}{12n_e}
        \frac{l_1l_2}{d^2}\right].
    \end{aligned}
\end{equation}
Taking the limit $n_e\rightarrow 1$ the first term vanishes
and we also get rid of the short-distance cutoff
\begin{equation}
    \begin{aligned}
        \mathcal{E}_{n_e\rightarrow 1}^{(2)} = 
        \frac{\chi_F}{4}
        \frac{l_1l_2}{d^2}.
    \end{aligned}
\end{equation}
Therefore it is proportional to the length of the two regions
and decays with the distance by a power law. 

\subsection{\label{hdfree}free fermion in higher dimension}
Similar to the 1+1-dimensional case, we consider free-fermion
system with conserved charge in $(D+1)$-dimension.
Here we consider region $A$ and $B$ to be two balls with 
radius $R_1, R_2$ and the distance between their centers $d\gg R_1, R_2$.
In higher dimensions the $M$-th order connected
charge correlator satisfies 
\begin{equation}
    \int \frac{d\bm{q}_M}{2\pi}\braket{\rho_{\bm{q}_1}
    \cdots\rho_{\bm{q}_M}}_c
    \sim k_F^{D+1-M} q^{M-1}
\end{equation}
for small $\bm{q}$ \cite{PhysRevX.12.031022},
where $q$ is the typical length of $\bm{q}_i$, $i=1,\cdots
M-1$.
In \eqref{nega_expansion_2_sec} the second order correlator
is
\begin{equation}
    \int \frac{d\bm{q}'}{2\pi}\braket{\rho_{\bm{q}}
    \rho_{\bm{q}'}}_c
    \sim k_F^{D-1} |\bm{q}|.
\end{equation}
by Fourier transformation it gives the real space correlator
\begin{equation}
    \braket{\rho(\bm{r}_1)\rho(\bm{r}_2)}_c
    \sim\frac{k_F^{D-1}}{|\bm{r}_1-\bm{r}_2|^{D+1}},
\end{equation}
for $|\bm{r}_1-\bm{r}_2|\gg\varepsilon$.
Taking the similar model for short distance cutoff
as the 1+1-d case, the integral in real space gives
\begin{equation}
    \begin{aligned}
        \braket{Q_A^2}_c 
        &\sim (k_FR_1)^{D-1}\left(
            1+O\left(\frac{R_1}{\varepsilon}\right)
        \right),
    \end{aligned}
\end{equation}
\begin{equation}
    \begin{aligned}
        \braket{Q_B^2}_c 
        &\sim (k_FR_2)^{D-1}\left(
            1+O\left(\frac{R_2}{\varepsilon}\right)
        \right),
    \end{aligned}
\end{equation}
and
\begin{equation}
    \begin{aligned}
        \braket{Q_AQ_B}_c 
        &\sim k_F^{D-1}\frac{R_1^DR_2^D}{d^{D+1}}.
    \end{aligned}
\end{equation}
Therefore \eqref{nega_expansion_2_sec} becomes
\begin{equation}
    \begin{aligned}
        \mathcal{E}_{n_e}^{(2)} 
        \sim&k_F^{D-1}\left[
        c_1\frac{n_e^2-1}{6n_e}\left(R_1^{D-1}+R_2^{D-1}
        +O\left(\frac{R}{\varepsilon}\right)\right)
        \right.\\
        &\left.+c_2\frac{n_e^2+2}{6n_e}
        \frac{R_1^DR_2^D}{d^{D+1}}
        \right],
    \end{aligned}
\end{equation}
where $c_1, c_2$ are some finite coefficient.
We see that when regions $A$ and $B$ are far apart, the terms from
$\braket{Q_A^2}_c$ and $\braket{Q_B^2}_c$
contain a constant term with short-distance cutoff that do not depend on the distance between 
$A$ and $B$, and is much larger than the terms from $\braket{Q_AQ_B}_c $. This incidates that R{\'e}nyi
negativity is not a good entanglement measure.
However, at replica limit $n_e\rightarrow 1$, the first term with short
distance cutoff
is also eliminated, and we get
\begin{equation}
    \begin{aligned}
        \mathcal{E}_{n_e\rightarrow 1}^{(2)} 
        \sim&k_F^{D-1}
        \frac{R_1^DR_2^D}{d^{D+1}}.
    \end{aligned}
\end{equation}
This is consistent with our expectation for logarithmic negativity as an 
entanglement measure.
The negativity is proportional to the area of two regions
and decays also by a power law with the distance between them.

\section{Conclusion and discussion}

In this paper, we derive the connected charge correlator expansion 
for R{\'e}nyi and logarithmic negativity in free-fermion systems 
with conserved $U(1)$ charge. We numerically check the validity of 
the expansion, and find that the convergence at replica limit requires
translational symmetry. Using this expansion, we analyze the 
scaling behavior of negativity in extensive free-fermion systems. 
From the expansion, it is direct to note that there are 
terms that diverge for short distance cutoff in the expansion for 
R{\'e}nyi negativity, but they vanish at the limit $n_e\rightarrow 1$.
This is consistent with our expectation for logarithmic negativity rather than 
R{\'e}nyi negativity as an 
entanglement measure.


Our research in charge correlator expansion is generally 
limited to free-fermion systems. One can also consider 
canonical bosonic systems where the charge operator is 
similarly defined, but will encounter difficulty in the 
range of $p$ \eqref{prange2}. For bosonic systems, 
the range of $p$ is given 
by $p = 0, 1, \cdots, n-1 \ (\mathrm{mod}\ n)$. There 
seems to be no natural choice of the range to make the 
summation $\sum_pp^M$ smallest for even $M$ and 
0 for odd $M$, as required for the convergence in fermionic systems. As a 
consequence, we cannot obtain the correct coefficient in 
the expansion after the terms with higher $M$ are truncated. 
Spin systems may also be considered, but the charge operator 
is hard to define, and free-spin systems (without spin-spin 
interaction) may not be intriguing for research.

The charge expansion may also be applied to other computable 
entanglement measures
besides PT negativity, \textit{e.g.} the computable cross norm or 
realignment (CCNR) negativity 
\cite{PhysRevA.67.032312,Rudolph2005,MR1985541}. 
Recently, it has been shown that R{\'e}nyi and logarithmic CCNR negativity 
of two intervals that are not adjacent
has a universal expression \cite{PhysRevLett.130.131601} in (1+1)-d CFT. 
The twist operators for CCNR negativity cannot be simultaneously 
diagonalized, but they can be simultaneously block 
diagonalized in the form of $2\times 2$ matrices. 
Therefore CCNR negativity can be in free systems can be related to
the expectation value of $2\times 2$ 
operators instead of $U(1)$ charge operators.
We suppose that the charge expansion for these computable entanglement 
measures can provide a deeper understanding of the 
entanglement properties of quantum systems.

Formula \eqref{univ_corr} shows that the universal $(D+1)$-th 
order charge correlator in $D$-dimensional free-fermion systems 
is proportional to the Euler characteristic of the Fermi sea. 
The leading term in the expansion for R{\'e}nyi and logarithmic negativity 
\eqref{nega_expansion_2_sec} and \eqref{log_nega_2} contains second-order charge 
correlators, and therefore in (1+1)-d system, it extracts the 
Euler characteristic, as shown in Sec. \ref{1dfree}. However, for 
higher dimensions, the divergent terms always 
exist in the leading term. It is an interesting question whether there are 
ways to subtract the divergent terms and extract topological terms in the expansion for negativity, 
like the case for topological entanglement entropy \cite{PhysRevX.12.031022}.

\begin{acknowledgments}
We thank Yingfei Gu for suggesting 
the topic, advising on various stages of this 
project, and comments on the draft.
We thank Pok Man Tam for detailed discussion on the derivation
and constructive suggestions. We thank Liang Mao, Zhenhuan Liu, Chao Yin, 
Yi-Fei Wang, Yuzhen Zhang and Zhenhua Zhu for 
valuable discussions.
\end{acknowledgments}

\bibliography{ref.bib}

\onecolumngrid

\appendix

\section{\label{appendix_entropy}Cumulant 
expansion for R{\'e}nyi and von Neumann entropy}
From \eqref{Tnexpanded}
\begin{equation}
    \begin{aligned}
        \log\braket{T_{n}} 
            &=\sum_{M=1}^{\infty}
                \left(\sum_p p^M\right)\frac{1}{M!}\left(
                    \frac{2\pi i}{n}
                \right)^M\braket{Q_A^M}_c,
    \end{aligned}
\end{equation}
the summation
of $p$ is
\begin{equation}
    C_{n,M} = \sum_{p=-\frac{n-1}{2}}^{\frac{n-1}{2}}p^M
\end{equation}
It can be written in the form 
of generalized harmonic number \cite{CHOI20112220}
\begin{equation}
    H_n^{(s)}(z) = \sum_{j=1}^{n}\frac{1}{(j+z)^s}
    \ \ \ \ (n\in\mathbb{Z}^+, s\in\mathbb{C},
        z\in\mathbb{C}\backslash\mathbb{Z}^-).
\end{equation}

First consider the case of even $n$, the result is
\begin{equation}
    \begin{aligned}
        \sum_{p=-\frac{n-1}{2}}^{\frac{n-1}{2}}
        p^M 
        &= \left[1+(-1)^M\right]\times
        \left[\left(\frac{1}{2}\right)^M
        +\left(\frac{3}{2}\right)^M
        +\cdots+\left(\frac{n-1}{2}\right)^M\right]\\
        &=\begin{cases}
            2H_{n/2}^{(-M)}\left(-\frac{1}{2}\right)\ &\mathrm{even}\ M\\
            0\ &\mathrm{odd}\ M.
        \end{cases}
    \end{aligned}
\end{equation}
For the range of arguments we consider here, the
generalized harmonic number is related to the 
generalized Riemann zeta function (this is generally 
not correct).
\begin{equation}
    \zeta(n, a)=\sum_{k=0}^{\infty}\frac{1}{(k+a)^n}
\end{equation} 
by 
\begin{equation}
    H_n^{(s)}(z)=
    \zeta(s, z+1)-\zeta(s, z+n+1)
\end{equation} 
and $\zeta(n)=\zeta(n, 1)$.  Thus
\begin{equation}
    \begin{aligned}
        \log \braket{T_n}
        &=
        \sum_{\substack{M=2\\
        \mathrm{even}\ M}}^{\infty}
            2H_{n/2}^{(-M)}\left(-\frac{1}{2}\right)
                \frac{1}{M!}\left(
                    \frac{2\pi i}{n}
                \right)^M\braket{Q_A^M}_c
        \\
        &=\sum_{\substack{M=2\\
        \mathrm{even}\ M}}^{\infty}
            2\left[\zeta\left(-M, \frac{1}{2}\right)
            -\zeta\left(-M, \frac{n+1}{2}\right)
            \right]
                \frac{1}{M!}\left(
                    \frac{2\pi i}{n}
                \right)^M\braket{Q_A^M}_c.
    \end{aligned}
\end{equation}
For even $M\geq 0$, $\zeta\left(-M, 1/2\right)=0$, 
we get 
\begin{equation}\label{eqnPTresult}
    \begin{aligned}
        \log \braket{T_n}
        &=
        -\sum_{\substack{M=2\\
        \mathrm{even}\ M}}^{\infty}
            2
            \zeta\left(-M, \frac{n+1}{2}\right)
                \frac{1}{M!}\left(
                    \frac{2\pi i}{n}
                \right)^M\braket{Q_A^M}_c.
    \end{aligned}
\end{equation}

For odd $n$, it is easy to check that we get
\begin{equation}
    \begin{aligned}
        \log \braket{T_n}
        &=
        \sum_{\substack{M=2\\
        \mathrm{even}\ M}}^{\infty}
            2H_{(n-1)/2}^{(-M)}\left(0\right)
                \frac{1}{M!}\left(
                    \frac{2\pi i}{n}
                \right)^M\braket{Q_A^M}_c
        \\
        &=\sum_{\substack{M=2\\
        \mathrm{even}\ M}}^{\infty}
            2\left[\zeta\left(-M, 1\right)
            -\zeta\left(-M, \frac{n+1}{2}\right)
            \right]
                \frac{1}{M!}\left(
                    \frac{2\pi i}{n}
                \right)^M\braket{Q_A^M}_c.
    \end{aligned}
\end{equation}
For even $M\geq 0$, $\zeta\left(-M, 1\right)=0$, 
thus we get the same result
\begin{equation}\label{eqnPTresult2}
    \begin{aligned}
        \log \braket{T_n}
        &=
        -\sum_{\substack{M=2\\
        \mathrm{even}\ M}}^{\infty}
            2
            \zeta\left(-M, \frac{n+1}{2}\right)
                \frac{1}{M!}\left(
                    \frac{2\pi i}{n}
                \right)^M\braket{Q_A^M}_c.
    \end{aligned}
\end{equation}
This gives the cumulant expansion for R{\'e}nyi entropy \eqref{expan_renyi}.

Since the right hand side of \eqref{eqnPTresult2} is real
for all positive integer $n$, we
now extend it from $n\in\mathbb{Z}^+$ to $n\in\mathbb{R}^+$ 
and calculate the von Neumann 
entropy by taking the $n\rightarrow 1$ limit. 
The analytic continuation can differ with
periodic functions and affect the result, but we assume
it is correct here. By using
\begin{equation}
    \frac{d}{dn}\zeta(k, n) = -k \zeta(k+1, n),
\end{equation}
we get
\begin{equation}
    \begin{aligned}
        S_A &= \lim_{n\rightarrow 1} \frac{\log \braket{T_n}}{1-n}
        =
        \sum_{\substack{M=2\\
        \mathrm{even}\ M}}^{\infty}
            2\left.
            \frac{d}{dn}
            \zeta\left(-M, \frac{n+1}{2}\right)\right|_{n=1}
                \frac{1}{M!}({2\pi i})^M\braket{Q_A^M}_c\\
        &=
        \sum_{\substack{M=2\\
        \mathrm{even}\ M}}^{\infty}
            M\left.
            \zeta\left(1-M, \frac{n+1}{2}\right)\right|_{n=1}
                \frac{1}{M!}({2\pi i})^M\braket{Q_A^M}_c.
    \end{aligned}
\end{equation}
And by using the relation
\begin{equation}
    \zeta(1-M) = 2(2\pi)^{-M}\cos(\pi M/2)(M-1)!\zeta(M),
\end{equation}
we get
\begin{equation}
    \begin{aligned}
        S_A &=\sum_{\substack{M=2\\
            \mathrm{even}\ M}}^{\infty}
                M\zeta\left(1-M, 1\right)
                \frac{1}{M!}({2\pi i})^M\braket{Q_A^M}_c\\
        &=\sum_{\substack{M=2\\
                \mathrm{even}\ M}}^{\infty}
                2\zeta(M) \braket{Q_A^M}_c.       
    \end{aligned}
\end{equation}
This is the cumulant expansion for von Neumann
entropy \eqref{expan_vonneumann}.

\end{document}